\documentclass[nonacm,sigplan]{acmart}
\usepackage{subfig}
\usepackage{graphicx}
\usepackage{subcaption}
\usepackage{caption}
\begin{document}

\title{Swapping-Centric Neural Recording Systems}

\author{Muhammed Ugur}
\affiliation{
  \institution{Yale University}
  \postcode{06511}
  \country{}
}
\author{\centerline{\mbox{Raghavendra Pradyumna Pothukuchi}}}
\affiliation{
  \institution{Yale University}
  \postcode{06511}
  \country{}
}
\author{Abhishek Bhattacharjee}
\affiliation{
  \institution{Yale University}
  \postcode{06511}
  \country{}
}

\vspace{2mm}

\begin{abstract}
Neural interfaces read the activity of biological neurons to help advance the neurosciences and offer treatment options for severe neurological diseases. The total number of neurons that are now being recorded using multi-electrode interfaces is doubling roughly every 4-6 years \cite{Stevenson2011}. However, processing this exponentially-growing data in real-time under strict power-constraints puts an exorbitant amount of pressure on both compute and storage within traditional neural recording systems. Existing systems deploy various accelerators for better performance-per-watt while also integrating NVMs for data querying and better treatment decisions. These accelerators have direct access to a limited amount of fast SRAM-based memory that is unable to manage the growing data rates. Swapping to the NVM becomes inevitable; however, naive approaches are unable to complete during the refractory period of a neuron -- i.e., a few milliseconds -- which disrupts timely disease treatment. We propose co-designing accelerators and storage, with swapping as a primary design goal, using theoretical and practical models of compute and storage respectively to overcome these limitations.
\end{abstract}

\maketitle 
\pagestyle{plain} 

\section{Background and Motivation}

Neural recording systems directly measure and process electrophysiological data from the nervous system in real-time. Applications running on these systems range from storing data for offline analysis to providing closed-loop electrical stimulation to suppress seizures in patients with refractory epilepsy. The efficacy of these systems are determined by the quality of signal and the total number of channels, i.e., independent streams of neural data, being recorded. This is because many neurological and psychiatric disorders are network-level, requiring detailed recordings of as many neural circuits as possible. This has led to the proliferation of large-scale multi-electrode probes that are designed for intracranial placement. As a result, the total number of neurons that are being recorded have doubled roughly every 4-6 years \cite{Stevenson2011}. Current interfaces, such as the Neuropixel or Neuralink's N1 implant, record 144 Mbps and 545 Mbps of neural data respectively, with the ultimate goal of reading Tbps to capture the entire brain's activity.

\begin{figure}[!htb]
  \centering
  \begin{minipage}[b]{0.20\textwidth}%
    \centering
    \includegraphics[width=\textwidth]{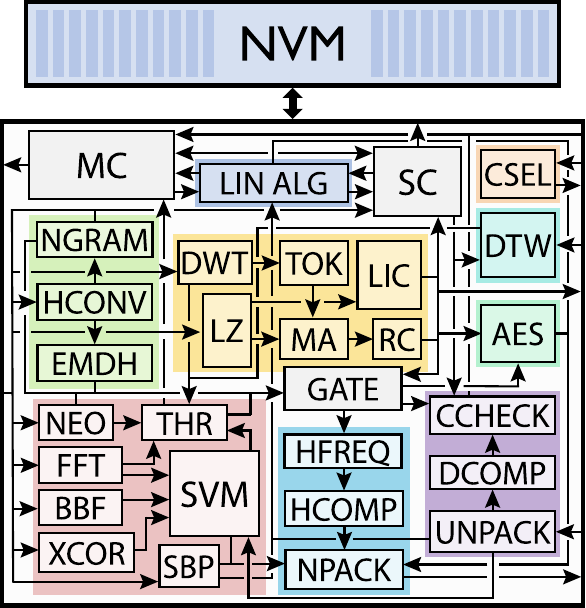}
  \end{minipage}
  \begin{minipage}[b]{0.25\textwidth}%
    \centering
    \includegraphics[width=0.85\textwidth]{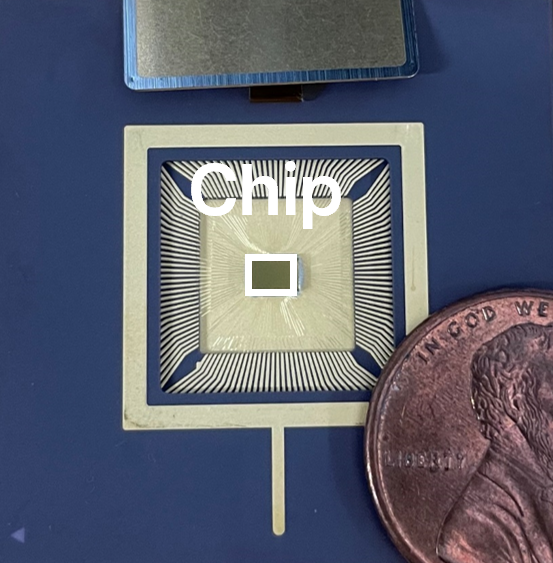}
  \end{minipage}
  \caption{Diagram for the most recent HALO processor ~\cite{Sriram2023}; composed of many unique accelerators, stitched together on a low-power reconfigurable fabric with access to a NVM (left). Partial tape-out of HALO at 12 nm (right).}
  \label{fig:processor}
\end{figure}

This exponential growth in data is pushing on-device neural signal processing to its limits, especially for invasive brain-computer interfaces (BCIs). Invasive interfaces possess up to two orders of magnitude better signal-to-noise ratio over non-invasive methods, providing more advanced treatment options. However, they must consume microwatts to milliwatts of power for safe operation, i.e., they must not heat more than $1^{\circ}$C of surrounding tissue. Staying within this low power regime while also supporting a variety of BCI applications has been accomplished with a heterogeneous reconfigurable array of course-grained accelerators in HALO (see Figure ~\ref{fig:processor}) \cite{karageorgos2020hardware}. However, these accelerators are implemented for a fixed number of channels at design-time and scale poorly as the number of channels increase. 

\begin{figure}[h]
\centering
\includegraphics[width=0.43\textwidth]{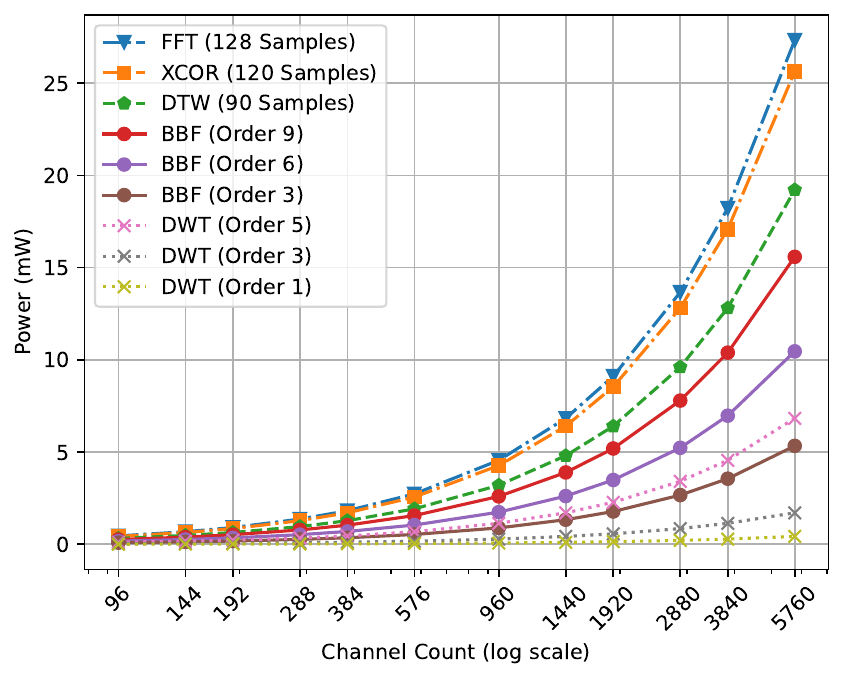}
\caption{Power usage of different accelerators when storing their working sets in SRAM as channel counts increase. Some accelerators, like the Butterworth Bandpass Filter (BBF) and Discrete Wavelet Transform (DWT), have configurations which vary their working sets. FFT, Cross Correlation (XCOR), and Dynamic Time Warping (DTW) are fixed at a set number of samples per channel. }
\label{fig:pe-power}
\end{figure}

Figure \ref{fig:pe-power} shows the power consumption of storing the working set of several signal processing accelerators in SRAM as the number of channels increase. Given the importance of power in this system, we rely on a detailed physical synthesis flow instead of standard architectural modeling. It only takes a single kernel, like the Fast Fourier Transform (FFT) at 128 samples, to overshoot a conservative 15 mW power budget as the number of channels scale. Data movement between accelerator memory and a better capacity-per-watt external storage becomes inescapable. NVM technologies, specifically NAND Flash, are a practical option and are already used in BCI applications for storing partial computations and raw signal data \cite{Sriram2023}. However, their bandwidth and latencies need to be considered carefully to meet real-time deadlines and stay within the power budget.

\section{System Design}
\begin{figure}[h]
\centering
\includegraphics[width=0.45\textwidth]{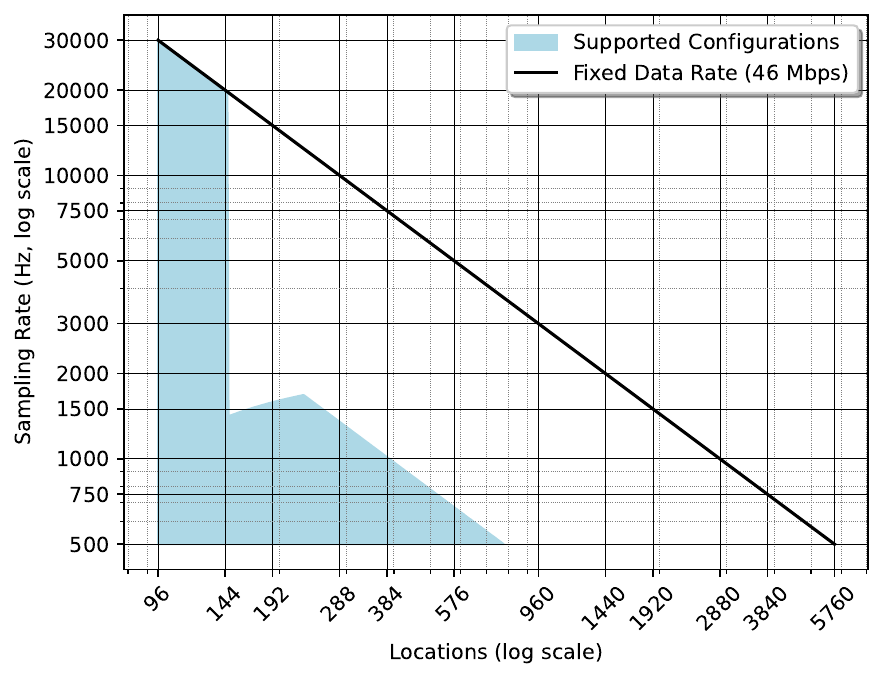}
\caption{Shows the support of channel/sampling-rate configurations (blue region) for a naive BBF swapping approach. Configurations that cannot fit into the SRAM of BBF and the storage controller require swapping. NVM bandwidth restricts higher channel counts whereas NVM write latency is too slow for smaller, yet uncacheable, working sets.}
\label{fig:channels}
\end{figure}

To better understand the impact of increasing channel counts on system resources, we fix the overall data rate of the system by reducing the sampling rate as channel counts grow. The total amount of samples that each accelerator processes in a given amount of time therefore remains constant across different channel/sampling-rate configurations. Figure ~\ref{fig:channels} shows the total amount of channel/sampling-rate configurations that can be supported for the Butterworth Bandpass Filter (BBF) on a 128GB Micron SLC NAND Flash chip \cite{micron}.

Naively moving data between the NVM and each of the accelerators significantly limits the amount of channels supported \textit{without increasing the incoming data rate}. These limitations are largely due to backend storage latency and bandwidth constraints. This analysis does not include power consumption, which would further reduce the amount of configurations supported. These constraints impose new design decisions for communicating with the NVM in these multi-accelerator systems to meet real-time deadlines. These challenges motivate the re-design of the accelerators and the storage sub-systems to be swapping-centric. Reducing the amount of data per I/O and the total number of I/Os per accelerator is therefore critical to support increasing channel counts at low-power. 

External memory algorithms are a class of techniques used to minimize the number of I/Os performed for an algorithm given some storage model \cite{Aggarwal1988}. These models however are outdated and do not consider EEPROMs like NAND Flash.  This work aims to incorporate these past techniques into the hardware-software co-design process by extending the algorithmic analysis to account for read/write performance asymmetry and power characteristics.  Prior external memory techniques also assume that the algorithm runs on a CPU with fixed main memory. Our approach proposes custom ASICs in hardware that implement the algorithms with the opportunity to add memory if necessary. This changes the compute/data movement tradeoffs across the memory hierarchy substantially and leads to unorthodox decisions at design-time. Finally, we have the opportunity to turn theory into hardware with this project by implementing our analysis decisions into the next generation of HALO processors. 

\section{NVM Modeling}
We use an abstract model for NAND Flash that takes into account data movement and power. This model includes the standard chip, die, plane, block, and page hierarchy, where multiple chips are connected together using a standard shared bus. Parallel operations exist across chips, dies, and planes, where chip-level parallelism costs more power due to turning on multiple chips at once. I/O costs include the data movement and memory operation latency/energy for page reads/programs. Global bandwidth is determined by bus characteristics, whereas chip-level bandwidth is determined by latency and/or simulation (e.g. using NVSim). General NVM specifications are gathered primarily through publicly available datasheets and other systems literature. 

\vspace{-1em}

\bibliographystyle{ACM-Reference-Format}
\bibliography{references}

\end{document}